\documentclass[aps,prl,twocolumn,showpacs,superscriptaddress,floatfix]{revtex4}
\def\FigFactor{0.5}

\usepackage{graphics}

\def\ET{\mbox{$E_T$}}
\def\pT{\mbox{$p_T$}}
\def\xT{\mbox{$x_T$}}

\def\sqrtsNN{\mbox{$\sqrt{s_\mathrm{_{NN}}}$}}
\def\sqrts{\mbox{$\sqrt{s}$}}
\def\Npart{\mbox{$\mathrm{N}_\mathrm{part}$}}
\def\NpartMean{\mbox{$\langle\Npart\rangle$}}
\def\Nbinary{\mbox{$\mathrm{N}_\mathrm{bin}$}}
\def\NbinaryMean{\mbox{$\langle\Nbinary\rangle$}}

\def\kzeros{\mbox{$K^0_s$}}
\def\pbar{\mbox{$\bar\mathrm{p}$}}
\def\lam{\mbox{$\Lambda$}}

\def\SigmaPlus{\mbox{$\Sigma^+$}}
\def\SigmaZero{\mbox{$\Sigma^0$}}

\def\pbarp{\mbox{$\pbar+\mathrm{p}$}}

\def\zvert{\mbox{$z_{vert}$}}
\def\zDCA{\mbox{$z_{DCA}$}}

\def\hplus{\mbox{$\mathrm{h}^+$}}
\def\hminus{\mbox{$\mathrm{h}^-$}}
\def\hphm{\mbox{$(\hplus+\hminus)/2$}}

\def\RAA{\mbox{$R_{AA}(\pT)$}}
\def\RCP{\mbox{$R_{CP}(\pT)$}}
\def\TAA{\mbox{$T_{AA}$}}

\def\Rsqrts{\mbox{$R_{200/130}(\pT)$}}

\def\sigmaNNinel{\mbox{$\sigma^{NN}_{inel}$}}
\def\sigmaAAgeom{\mbox{$\sigma^{AuAu}_{geom}$}}

\def\lt{\mbox{$<$}}
\def\gt{\mbox{$>$}}

\begin{document}

\title{
Transverse momentum and collision energy dependence of high \pT\
hadron suppression in Au+Au collisions at ultrarelativistic energies\\
}

%-----------------------------------------------------------------------------------------------
\affiliation{Argonne National Laboratory, Argonne, Illinois 60439}
\affiliation{Brookhaven National Laboratory, Upton, New York 11973}
\affiliation{University of Birmingham, Birmingham, United Kingdom}
\affiliation{University of California, Berkeley, California 94720}
\affiliation{University of California, Davis, California 95616}
\affiliation{University of California, Los Angeles, California 90095}
\affiliation{Carnegie Mellon University, Pittsburgh, Pennsylvania 15213}
\affiliation{Creighton University, Omaha, Nebraska 68178}
\affiliation{Nuclear Physics Institute AS CR, \v{R}e\v{z}/Prague, Czech Republic}
\affiliation{Laboratory for High Energy (JINR), Dubna, Russia}
\affiliation{Particle Physics Laboratory (JINR), Dubna, Russia}
\affiliation{University of Frankfurt, Frankfurt, Germany}
\affiliation{Indiana University, Bloomington, Indiana 47408}
\affiliation{Insitute  of Physics, Bhubaneswar 751005, India}
\affiliation{Institut de Recherches Subatomiques, Strasbourg, France}
\affiliation{University of Jammu, Jammu 180001, India}
\affiliation{Kent State University, Kent, Ohio 44242}
\affiliation{Lawrence Berkeley National Laboratory, Berkeley, California 94720}
\affiliation{Max-Planck-Institut f\"ur Physik, Munich, Germany}
\affiliation{Michigan State University, East Lansing, Michigan 48824}
\affiliation{Moscow Engineering Physics Institute, Moscow Russia}
\affiliation{City College of New York, New York City, New York 10031}
\affiliation{NIKHEF, Amsterdam, The Netherlands}
\affiliation{Ohio State University, Columbus, Ohio 43210}
\affiliation{Panjab University, Chandigarh 160014, India}
\affiliation{Pennsylvania State University, University Park, Pennsylvania 16802}
\affiliation{Institute of High Energy Physics, Protvino, Russia}
\affiliation{Purdue University, West Lafayette, Indiana 47907}
\affiliation{University of Rajasthan, Jaipur 302004, India}
\affiliation{Rice University, Houston, Texas 77251}
\affiliation{Universidade de Sao Paulo, Sao Paulo, Brazil}
\affiliation{University of Science \& Technology of China, Anhui 230027, China}
\affiliation{Shanghai Institute of Nuclear Research, Shanghai 201800, P.R. China}
\affiliation{SUBATECH, Nantes, France}
\affiliation{Texas A \& M, College Station, Texas 77843}
\affiliation{University of Texas, Austin, Texas 78712}
\affiliation{Valparaiso University, Valparaiso, Indiana 46383}
\affiliation{Variable Energy Cyclotron Centre, Kolkata 700064, India}
\affiliation{Warsaw University of Technology, Warsaw, Poland}
\affiliation{University of Washington, Seattle, Washington 98195}
\affiliation{Wayne State University, Detroit, Michigan 48201}
\affiliation{Institute of Particle Physics, CCNU (HZNU), Wuhan, 430079 China}
\affiliation{Yale University, New Haven, Connecticut 06520}
\affiliation{University of Zagreb, Zagreb, HR-10002, Croatia}

%Authors 1 per line follow by affiliation, three authors have two affiliations
\author{J.~Adams}\affiliation{University of Birmingham, Birmingham, United Kingdom}
\author{C.~Adler}\affiliation{University of Frankfurt, Frankfurt, Germany}
\author{M.M.~Aggarwal}\affiliation{Panjab University, Chandigarh 160014, India}
\author{Z.~Ahammed}\affiliation{Purdue University, West Lafayette, Indiana 47907}
\author{J.~Amonett}\affiliation{Kent State University, Kent, Ohio 44242}
\author{B.D.~Anderson}\affiliation{Kent State University, Kent, Ohio 44242}
\author{M.~Anderson}\affiliation{University of California, Davis, California 95616}
\author{D.~Arkhipkin}\affiliation{Particle Physics Laboratory (JINR), Dubna, Russia}
\author{G.S.~Averichev}\affiliation{Laboratory for High Energy (JINR), Dubna, Russia}
\author{S.K.~Badyal}\affiliation{University of Jammu, Jammu 180001, India}
% ...
\collaboration{STAR Collaboration}\homepage{www.star.bnl.gov}\noaffiliation

\date{\today}

%-----------------------------------------------------------------------------------------------

\begin{abstract}
We report high statistics measurements
of inclusive charged hadron production in Au+Au and p+p collisions at
\sqrtsNN=200 GeV. A large, approximately constant hadron suppression 
is observed in central Au+Au collisions for $5\lt\pT\lt12$ GeV/c. The
collision energy dependence of the yields and the centrality and \pT\
dependence of the suppression provide stringent constraints on
theoretical models of suppression. Models incorporating initial-state
gluon saturation or partonic energy loss in dense matter are largely
consistent with observations. We observe no evidence of
\pT-dependent suppression, which may be expected from models
incorporating jet attentuation in cold nuclear matter or scattering of
fragmentation hadrons.

\end{abstract}

\pacs{25.75.Dw,25.75.-q,13.85.Hd}

\maketitle

%=================================================================================
% Intro
%=================================================================================

%PARA1
High energy partons propagating through matter are predicted to lose
energy via induced gluon radiation, with the total energy loss
strongly dependent on the color charge density of the medium
\cite{EnergyLoss}. This process can provide a sensitive probe of the
hot and dense matter generated early in ultrarelativistic nuclear
collisions, when a plasma of deconfined quarks and gluons may
form. The hard scattering and subsequent fragmentation of partons
generates jets of correlated hadrons. In nuclear collisions, jets may be studied 
via observables such as high transverse momentum (high \pT) hadronic inclusive 
spectra \cite{WangGyulassyLeadinHadron} and correlations. Several striking
high \pT\ phenomena have been observed at the Relativistic Heavy Ion
Collider (RHIC)
\cite{STARHighptSpectra,PHENIXhighpt,PHOBOSNpartScaling,STARHighptFlow,STARHighptCorrelations},
including strong suppression of inclusive hadron production
\cite{STARHighptSpectra,PHENIXhighpt,PHOBOSNpartScaling}.  
These phenomena are consistent with large partonic energy loss in high
energy density matter 
\cite{EnergyLoss,Suppression,VitevGyulassy,SalgadoWiedemann,HiranoNara}
though other mechanisms have been proposed, including gluon saturation 
in the initial nuclear wavefunction \cite{KharzeevLevinMcLerran}, attenuation of 
jet formation in cold nuclear matter \cite{TomasikPartonFormationTime},
and scattering of fragmentation hadrons \cite{GallmeisterEtAl}.  Additional 
measurements are required to discriminate among these pictures and to isolate 
effects due to final state partonic energy loss.

%PARA2
We report high statistics measurements by the STAR Collaboration 
of the inclusive charged hadron yield \hphm\ (defined as the summed yields of 
primary $\pi^\pm$, K$^\pm$, p and \pbar) in Au+Au collisions and non-singly 
diffractive (NSD) p+p collisions at nucleon-nucleon center of mass energy 
\sqrtsNN=200 GeV. The Au+Au data extend considerably the \pT\ range of earlier 
charged hadron suppression studies, and the p+p data are the first such 
measurement at this energy. Comparisons are made to several theoretical models. 
The high precision and broad kinematic coverage of the data significantly
constrain the possible mechanisms of hadron suppression. In addition,
the energy dependence of the yields may be sensitive to gluon
shadowing at low Bjorken $x$ in heavy nuclei.

%PARA3
We compare the data to two calculations based on hard parton
scattering evaluated via perturbative QCD (pQCD-I
\cite{XNWangPrivateComm} and pQCD-II \cite{VitevGyulassy}) and to a
calculation extending the saturation model to high momentum transfer
\cite{KharzeevLevinMcLerran}. Both pQCD models for Au+Au collisions
incorporate nuclear shadowing, the Cronin effect \cite{CroninEffect}, 
and partonic energy loss, but with different formulations. pQCD-I results 
excluding one or more nuclear effects are also shown.  Neither pQCD calculation 
includes non-perturbative effects that generate particle species-dependent 
differences for \pT\lt5 
GeV/c\cite{VitevGyulassyBaryonEnhancement,XNWangPrivateComm}.

%=================================================================================
% STAR technical
%=================================================================================

%PARA4
Charged particle trajectories were measured in the Time Projection Chamber (TPC) 
\cite{TPC}.  The magnetic field was 0.5 T, resulting in a factor of three 
improvement in \pT\ resolution at high \pT\ relative to 
\cite{STARHighptSpectra,STARHighptFlow}.  After event selection cuts, the Au+Au 
dataset comprised 1.7 million minimum bias events ($97\pm3\%$ of the geometric 
cross section \sigmaAAgeom) and 1.5 million central events (10\% of
\sigmaAAgeom). Centrality selection and analysis of spectra follow
Ref.~\cite{STARHighptSpectra}. Background at high \pT\ is dominated by weak decay 
products, with correction factors calculated using STAR measurements of 
\lam(+\SigmaZero) and \kzeros\ for \pT\lt6 GeV/c \cite{STARLambdaK0sToBePublished} 
and assuming constant yield ratios \lam(+\SigmaZero)/(\hplus+\hminus) and 
\kzeros/(\hplus+\hminus) for \pT\gt6 GeV/c. The \lam(+\SigmaZero) yield was scaled 
by a factor 1.4 to account for \SigmaPlus\ decays.  Table \ref{TableCorr} 
summarizes the correction factors at high \pT.

%------------------------------------------------------------------------------------------------------
\begin{table}
\caption{Multiplicative correction factors applied to the measured yields at 
\pT=10 GeV/c for p+p and Au+Au data. Factors vary by approximately 5\% within
4\lt\pT\lt12 GeV/c and have similar uncertainties.
\label{TableCorr}}
\begin{ruledtabular}
\begin{tabular}{c||c|c|c}
 & Tracking & Background & \pT\ resolution \\
\hline
p+p & 1.18 $\pm$ 0.07 & 0.90 $\pm$ 0.08  & 0.89 $^{+0.05}_{-0.05}$ \\
Au+Au 60-80\% & 1.11 $\pm$ 0.06 & 0.95 $\pm$ 0.05 & 0.97 $^{+0.03}_{-0.05}$ \\
Au+Au 0-5\% & 1.25 $\pm$ 0.06 & 0.94 $\pm$ 0.06  & 0.95 $^{+0.05}_{-0.05}$ \\
\end{tabular}
\end{ruledtabular}
\end{table}
%-------------------------------------------------------------------------

%PARA5
After event selection cuts, the p+p dataset comprised 5 million mainly
NSD events, triggered on the coincidence of two Beam-Beam counters
(BBCs). The BBCs are annular scintillator detectors situated $\pm$3.5
m from the interaction region, covering pseudorapidity
3.3\lt$|\eta|$\lt5.0. A van der Meer scan \cite{VernierScan} measured
the BBC trigger cross section to be 26.1$\pm$0.2(stat)$\pm$1.8(sys)
mb. The BBC trigger was simulated using PYTHIA \cite{PYTHIA} and
HERWIG \cite{HERWIG} events passed though a GEANT detector model. The
PYTHIA trigger cross section is 27 mb, consistent with measurement, of
which 0.7 mb results from singly diffractive events. The PYTHIA and HERWIG
simulations show that the trigger accepts 87$\pm$8\% of all NSD events
containing a TPC track, with negligible track \pT-dependence. Non-interaction 
backgrounds contributed 3$\pm$2\% of the trigger rate.  The high p+p interaction 
rate generated significant pileup in the TPC. Valid tracks matched an in-time hit 
in the Central Trigger Barrel (CTB \cite{TPC}) surrounding the TPC and projected 
to a distance of closest approach $DCA$\lt1 cm to the average beam trajectory. To 
avoid event selection multiplicity bias, an approximate event vertex position 
along the beam (\zvert) was calculated by averaging \zDCA\ over all valid tracks. 
Accepted events were required to have $|\zvert|$\lt75 cm, corresponding to 
69$\pm$4\% of all events. The track \pT\ fit did not include the event vertex. The 
CTB track-matching efficiency is 94$\pm$2\% and combinatorial background is 
2$\pm$2\%. Other significant p+p tracking backgrounds result from weak decays and
antinucleon annihilation in detector material, with corrections calculated using 
HIJING \cite{Hijing} and preliminary STAR measurements. Correction factors at high 
\pT\ are given in Table~\ref{TableCorr}. For p+p collisions relative to peripheral
Au+Au, exclusion of the event vertex from the \pT\ fit results in poorer \pT\ 
resolution, while the CTB matching requirement results in lower tracking 
efficiency. The p+p inclusive spectrum was also analysed for \pT\lt3.5 GeV/c by an 
independent method in which a primary vertex is found and incorporated into the 
track fit, with consistent results.

%-----------------------------------------------------------------------
\begin{figure}
\resizebox{.5\textwidth}{!}{\includegraphics{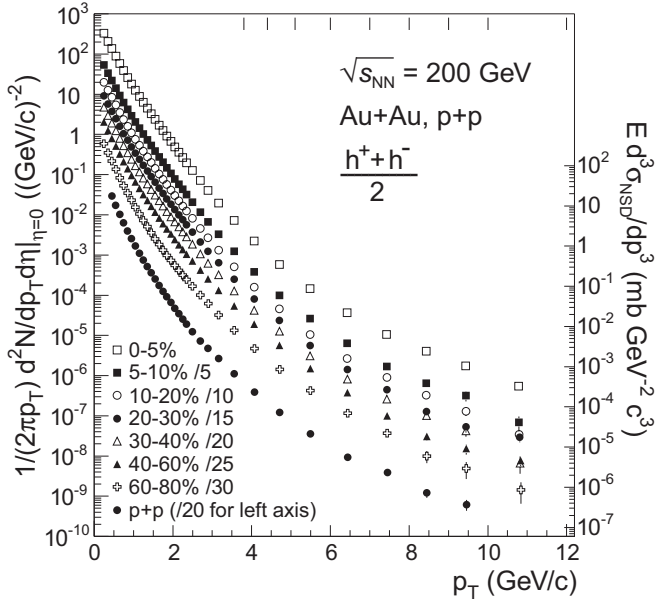}}
\caption{Inclusive invariant \pT\ distributions of \hphm\ for centrality-selected 
Au+Au and p+p NSD interactions. Hash marks at the top indicate bin boundaries for 
\pT\gt4 GeV/c. The invariant cross section for p+p is indicated on the right 
vertical axis.
\label{FigOne}}
\end{figure}
%-----------------------------------------------------------------------

%PARA6
The p+p NSD differential cross section is the product of the measured per-event 
yield and the BBC NSD trigger cross section, and has a normalization uncertainty 
of $\pm$14\%. The charged hadron invariant cross section has been measured in 
\pbarp\ collisions at \sqrts=200 GeV \cite{UA1}. The p+p cross section reported 
here is smaller by a factor of $0.79\pm0.18$, approximately independent of \pT, 
where the uncertainty includes the two spectrum normalizations and the correction 
for different acceptances \cite{STARHighptSpectra}. The difference is due in large 
part to differing NSD cross section, which is $35\pm1$ mb in \cite{UA1} but is 
measured here to be $30.0\pm3.5$ mb.

%=================================================================================
% Inclusive spectra
%=================================================================================

%PARA7
Figure~\ref{FigOne} shows inclusive invariant \pT\ distributions of \hphm\ within 
$|\eta|$\lt0.5 for Au+Au and p+p collisions at \sqrtsNN=200 GeV. The Au+Au spectra 
are shown for percentiles of \sigmaAAgeom, with 0-5\% indicating the most central 
(head-on) collisions.  Error bars are the quadrature sum of the statistical and 
systematic uncertainties and are dominated by the latter except at the highest \pT.

%=================================================================================
% 200/130 Ratios of spectra
%=================================================================================

\begin{figure}
\resizebox{\FigFactor\textwidth}{!}{\includegraphics{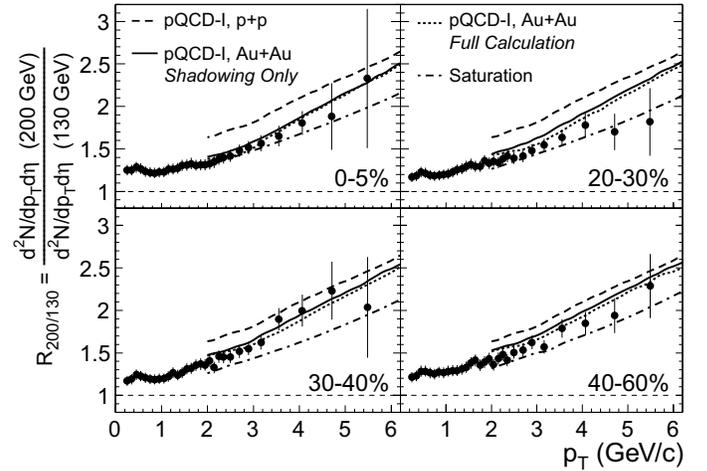}}
\caption{\Rsqrts\ vs. \pT\ for \hphm\ for four different centrality bins.  The 
overall normalization uncertainty is $^{+6}_{-10}$\% for the 40-60\% bin and is 
negligible for the other panels. Calculations are described in the text.}
\label{FigTwo}
\end{figure}

%PARA8
Figure \ref{FigTwo} shows \Rsqrts, the ratio of charged hadron yields at
\sqrtsNN=200 and 130 GeV\cite{STARHighptSpectra}, for centrality selected Au+Au 
collisions.  Error bars are the quadrature sum of the statistical and systematic
uncertainties, dominated for \pT\gt4 GeV/c by statistics at 130 GeV. In the 
absence of nuclear effects, the hard process inclusive yield in nuclear collisions 
is expected to scale as \NbinaryMean, the average number of binary collisions for 
the respective centrality selection. \Rsqrts\ has not been scaled by the ratio
\Nbinary(200)/\Nbinary(130), which Glauber model calculations 
\cite{STARHighptSpectra,Glauber} give as $\sim1.02$ for all centralities.  
Figure~\ref{FigTwo} also shows the saturation model calculation and pQCD-I 
calculations for p+p and centrality-selected Au+Au collisions (shadowing-only and 
full).  Both models approximately reproduce the \pT-dependence of the ratio for 
Au+Au for \pT\gt2 GeV/c, with pQCD-I slightly better for more peripheral 
collisions. The various pQCD-I calculations shown illustrate that in this model 
the reduction in \Rsqrts\ for Au+Au relative to p+p is predominantly due to 
nuclear shadowing\cite{XNWangPrivateComm}. This sensitivity arises because the
shadowing is $x$-dependent and at fixed \pT, different \sqrts\ corresponds to 
different $\xT=2\pT/\sqrts$. The quantitative agreement of pQCD-I with the data 
improves for more peripheral collisions, suggesting that the prescription for the 
centrality dependence of shadowing in \cite{XNWangPrivateComm} may not be optimal. 
Alternatively, introduction of \sqrts-dependent energy loss to the model in 
\cite{XNWangPrivateComm} may also improve the agreement.

%-------------------------------------------------------------------------
%-------------------------------------------------------------------------

%=================================================================================
% STAR RAA
%=================================================================================
 
%PARA9
Nuclear effects on the inclusive spectra are measured by comparison to a 
nucleon-nucleon (NN) reference via the nuclear modification factor:
\begin{equation}
\label{RAA}
\RAA=\frac{d^2N^{AA}/d{\pT}d\eta}{\TAA{d}^2\sigma^{NN}/d{\pT}d{\eta}}\ ,
\end{equation}
\noindent
where \TAA=\NbinaryMean/\sigmaNNinel\ from a Glauber calculation accounts for the 
nuclear collision geometry \cite{STARHighptSpectra,Glauber} and we adopt 
\sigmaNNinel\ = 42 mb. ${d}^2\sigma^{NN}/d{\pT}d{\eta}$ refers to inelastic 
collisions, whereas we have measured the p+p NSD differential cross section. 
However, singly diffractive interactions contribute predominantly to low \pT\ 
\cite{ZeusDiffraction}. A multiplicative correction based on PYTHIA, applied to
${d}^2\sigma^{NN}/d{\pT}d{\eta}$ in Eq.~\ref{RAA}, is 1.05 at \pT=0.4 GeV/c and 
unity above 1.2 GeV/c.

\begin{figure}
\resizebox{\FigFactor\textwidth}{!}{\includegraphics{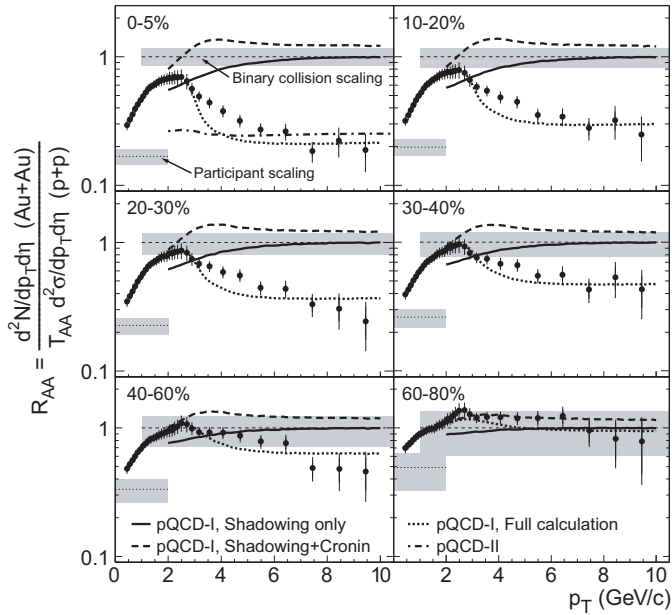}}
\caption{\RAA\ (Eq.~\ref{RAA}) for \hphm\ in $|\eta|$\lt0.5, for 
centrality-selected Au+Au spectra relative to the measured p+p spectrum. The p+p 
spectrum is common to all panels. Calculations are described in the text.
\label{FigThree}}
\end{figure}

%PARA10
Figure~\ref{FigThree} shows \RAA\ at \sqrtsNN=200 GeV for centrality-selected 
Au+Au spectra relative to the measured p+p spectrum. Horizontal dashed lines show 
Glauber model expectations \cite{STARHighptSpectra,Glauber} for scaling of the 
yield with \NbinaryMean\ or mean number of participants \NpartMean, with grey 
bands showing their respective uncertainties summed in quadrature with the p+p 
normalization uncertainty. The error bars represent the quadrature sum of the 
Au+Au and remaining p+p spectrum uncertainties. For \pT\lt6 GeV/c, \RAA\ is 
similar to that observed at \sqrtsNN=130 GeV \cite{STARHighptSpectra}.  Hadron 
production for 6\lt\pT\lt10 GeV/c is suppressed by a factor of 4-5 in central 
Au+Au relative to p+p collisions.

%PARA11
Figure~\ref{FigThree} also shows the full pQCD-I calculation and the influence of 
each nuclear effect. The energy loss for central collisions is a fit parameter, 
with the \pT\ and centrality dependence of the suppression constrained by theory. 
The Cronin enhancement and shadowing alone cannot account for the suppression, 
which is reproduced only if partonic energy loss in dense matter is included. The 
full calculation generally agrees with data for \pT\gt5 GeV/c if the initial 
parton density in central collisions is adjusted to be $\sim$15 times that of cold 
nuclear matter \cite{dEdxColdMatter}. pQCD-II exhibits similar agreement for
central collisions. In Ref.~\cite{VitevGyulassy}, the pQCD-II calculation was used 
to predict a \pT-independent suppression factor in this \pT\ range from the 
interplay between shadowing, the Cronin effect, and partonic energy loss.

%-----------------------------------------------------------------------
%----------------------------------------------------------------------

%=================================================================================
% Central Over Peripheral
%=================================================================================

\begin{figure}
\resizebox{\FigFactor\textwidth}{!}{\includegraphics{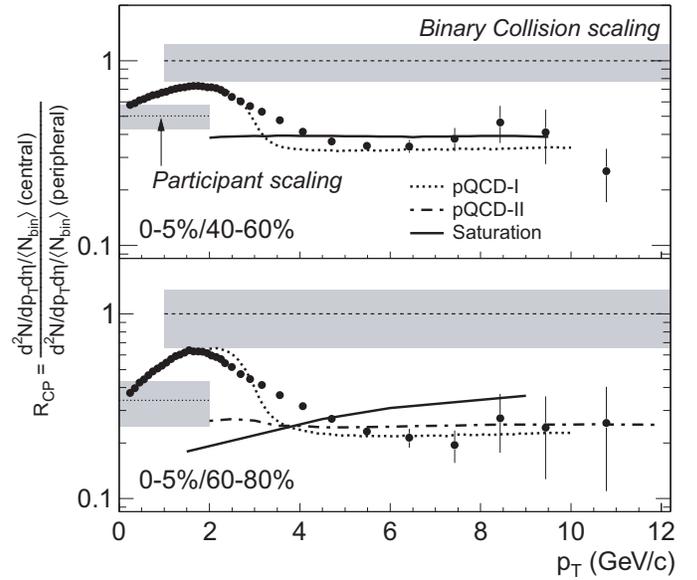}}
\caption{\RCP\  vs. \pT\ for \hphm.  Calculations are described in the text.
\label{FigFour}}
\end{figure}

%PARA12
Figure~\ref{FigFour} shows \RCP, the \NbinaryMean-normalized ratio of central and 
peripheral Au+Au spectra. \RCP\ extends to higher \pT\ than \RAA, with smaller 
point-to-point uncertainties. The error bars show the quadrature sum of 
statistical and uncorrelated systematic uncertainties. Statistical errors dominate
the uncertainties for \pT\gt8 GeV/c.  Lines and grey bands are as in 
Fig.~\ref{FigThree}.  \RCP\ for \pT\lt6 GeV/c is similar to measurements at 
\sqrtsNN=130 GeV \cite{STARHighptSpectra}, but is now seen to be approximately
constant for 5\lt\pT\lt12 GeV/c.  It is consistent with \NpartMean\ scaling at 
$\pT\sim4$ GeV/c as reported in \cite{PHOBOSNpartScaling}, but is below 
\NpartMean\ scaling at higher \pT.

%-----------------------------------------------------------------------
%-----------------------------------------------------------------------

%PARA13
The \pT-dependence of the suppression in Figure \ref{FigFour} is well reproduced 
for \pT\gt5 GeV/c by the full pQCD-I and pQCD-II calculations in both panels and 
the saturation calculation in the upper but not the lower panel. The magnitude of 
suppression is fitted to the central collision data in the pQCD models but is 
predicted in the saturation calculation.  Attenuation of initial jet formation due 
to multiple nucleon interactions \cite{TomasikPartonFormationTime} generates an 
increase in partonic \RAA\ for central collisions of a factor $\sim2$ in 
$5\lt\ET\lt12$ GeV.  A similar \pT-dependence would be expected for high \pT\ 
hadrons, in contrast to observations. Suppression in the final state due to 
in-medium scattering of fragmentation hadrons should also result in a rising \RAA\ 
with increasing \pT\ due to the dependence of hadron formation time on the total 
jet energy \cite{GallmeisterEtAl}, though detailed comparison of this model to 
data requires further theoretical development.

%=================================================================================
% Conclusions
%=================================================================================

%PARA14
In summary, STAR has measured inclusive charged hadron yields from Au+Au and p+p 
collisions at \sqrtsNN=200 GeV, at higher precision and over a much broader \pT\ 
range than previous measurements. Large, constant hadron suppression is observed 
in central nuclear collisions at high \pT. The systematic behaviour of the 
suppression at high \pT\ is well described both by pQCD calculations incorporating 
final-state partonic energy loss in dense matter and a model of initial-state
gluon saturation, though the latter model provides a poorer description of 
peripheral collision data. The isolation of initial state effects on high \pT\ 
hadron production may be achieved through the study of d+Au collisions at RHIC, 
allowing a quantitative measurement of final state effects from the data presented 
here.

%============================================================================

\begin{acknowledgments}
We thank C.~Greiner, M.~Gyulassy, D.~Kharzeev, C.~Salgado, B.~Tomasik,
I.~Vitev and X.N.~Wang for valuable discussions.  
We thank the RHIC Operations Group and RCF at BNL, and the
NERSC Center at LBNL for their support. This work was supported
in part by the HENP Divisions of the Office of Science of the U.S.
DOE; the U.S. NSF; the BMBF of Germany; IN2P3, RA, RPL, and
EMN of France; EPSRC of the United Kingdom; FAPESP of Brazil;
the Russian Ministry of Science and Technology; the Ministry of
Education and the NNSFC of China; SFOM of the Czech Republic,
DAE, DST, and CSIR of the Government of India; the Swiss NSF.
\end{acknowledgments}


\def\etal{\mbox{$\mathrm{\it et\ al.}$}}
\def\QM{Proceedings of Quark Matter 2001, Stony Brook, N.Y.,
Jan. 15-21, 2001. Nucl.\ Phys.\ A, in print.}
\def\QM2002{Proceedings of Quark Matter 2002, Nantes, France,
Jul. 18-24, 2002. Nucl.\ Phys.\ A, to be published.}


\begin{thebibliography}{99}

\bibitem{EnergyLoss} R.~Baier, D.~Schiff and B.G.~Zakharov, Ann. Rev. Nucl. Part. Sci. {\bf 50}, 37 (2000); 
M.~Gyulassy, I.~Vitev, X.N.~Wang, B.~Zhang, nucl-th/0302077.

\bibitem{WangGyulassyLeadinHadron} X.N.~Wang and M.~Gyulassy, Phys. Rev. Lett. {\bf 68}, 1480 (1992).

%--- these are refered to as a group in intro, don't change their order

\bibitem{STARHighptSpectra} C.~Adler \etal, Phys. Rev. Lett. {\bf 89}, 202301 (2002).  

\bibitem{PHENIXhighpt} K.~Adcox \etal, Phys. Rev. Lett. {\bf 88}; 022301 (2002); Phys. Lett. {\bf B561}, 82 (2003); 
S.S.~Adler \etal, Phys. Rev. Lett. {\bf 91}, 072301 (2003).

\bibitem{PHOBOSNpartScaling} B.B.~Back \etal, nucl-ex/0302015.

\bibitem{STARHighptFlow} C.~Adler \etal, Phys. Rev. Lett. {\bf 90}, 032301 (2003).

\bibitem{STARHighptCorrelations} C.~Adler \etal, Phys. Rev. Lett. {\bf 90}, 082302 (2003).

%--- end of group

\bibitem{Suppression} M.~Gyulassy and X.N.~Wang, Nucl. Phys. {\bf B420}, 583 (1994);
X.N.~Wang, Phys. Rev. {\bf C58}, 2321 (1998).

\bibitem{VitevGyulassy} I.~Vitev and M.~Gyulassy, Phys. Rev. Lett. {\bf 89}, 252301 (2002).

\bibitem{SalgadoWiedemann} C.A.~Salgado and U.A.~Wiedemann, Phys. Rev. Lett. {\bf 89}, 092303 (2002).

\bibitem{HiranoNara} T.~Hirano and Y.~Nara, nucl-th/0301042.

\bibitem{KharzeevLevinMcLerran} D.~Kharzeev, E.~Levin and L.~McLerran, Phys. Lett. {\bf B561}, 93 (2003);
D.~Kharzeev, private communication.

\bibitem{TomasikPartonFormationTime} R.~Lietava, J.~Pisut, N.~Pisutova, B.~Tomasik, 
Eur. Phys. J. {\bf C28}, 119 (2003).

\bibitem{GallmeisterEtAl} K.~Gallmeister, C.~Greiner and Z.~Xu, Phys. Rev. {\bf C67}, 044905 (2003). 
Note that Eq. (2) of this reference implies that
$\langle{L}/\lambda\rangle$ in Fig. 9 decreases substantially in
$5<\pT<12$ GeV/c.

\bibitem{XNWangPrivateComm} X.N.~Wang, nucl-th/0305010; private communication. Calculations 
use model parameters $\mu_0=2.0$ GeV and $\epsilon_0=2.04$ GeV/fm.

\bibitem{CroninEffect} D.~Antreasyan \etal, Phys. Rev. {\bf D19}, 764 (1979); 
P.B.~Straub \etal, Phys. Rev. Lett. {\bf 68}, 452 (1992).
 
\bibitem{VitevGyulassyBaryonEnhancement} I.~Vitev and M.~Gyulassy, Phys. Rev. {\bf 
C65}, 041902 (2002); R.J.~Fries, B.~Muller, C.~Nonaka and S.A.~Bass, Phys. Rev. Lett. {\bf 90} 202303 (2003).

\bibitem{TPC} M.~Anderson \etal, Nucl. Instr. Meth. {\bf A499}, 659 (2003).

\bibitem{STARLambdaK0sToBePublished} J.~Adams \etal, nucl-ex/0306007.

\bibitem{VernierScan} A.~Drees and Z.~Xu, Proceedings of the Particle Accelerator Conference 2001, Chicago, Il, p.
3120.

\bibitem{PYTHIA} T.~Sjoestrand \etal, Comp. Phys. Comm. {\bf 135}, 238 (2001).

\bibitem{HERWIG} G.~Corcella \etal, JHEP {\bf 0101}, 010 (2001).

\bibitem{Hijing} X.N.~Wang and M.~Gyulassy, Phys. Rev. {\bf D44}, 3501 (1991).

\bibitem{UA1} C.~Albajar \etal, Nucl. Phys. \textbf{B335}, 261 (1990).

\bibitem{Glauber} Correction for a recently found error in the Glauber calculation reported in
\cite{STARHighptSpectra} yields $\sim$2-6\% larger \NbinaryMean\ values, with the largest
increase for the most peripheral and central bins. Negligible change results
for \NpartMean. We thank K. Reygers for valuable discussions.

\bibitem{ZeusDiffraction} M.~Derrick \etal, 
Z. Phys. {\bf C67}, 227 (1995); R.E.~Ansorge \etal, Z. Phys. {\bf C33}, 175 (1986).

\bibitem{dEdxColdMatter} E.~Wang and X.N.~Wang, Phys. Rev. Lett. {\bf89}, 162301 (2002); F.~Arleo, Phys. Lett. {\bf 
B532}, 231 (2002).

%\bibitem{PHENIXBaryonEnhancement} J.~Velkovska, Proceedings of Strange Quark Matter 2003, J.~Phys.~G. to be published.

\end{thebibliography}

%============================================================================
\end{document}